\documentclass[12pt,a4paper]{article} 
\pdfoutput=1
\usepackage{jheppub}
\notoc

\usepackage{graphicx} 
\usepackage{amsmath}
\usepackage{xcolor}
\usepackage{soul}

\usepackage{cleveref}
\crefname{table}{Table}{Tables}
\crefname{equation}{Eq.}{Eqs.}
\crefname{appendix}{App.}{Apps.}
\crefname{section}{Sec.}{Secs.}
\crefname{figure}{Fig.}{Figs.}

\newcommand{\s}{\hspace{0.8pt}}

\definecolor{colorTC}{rgb}{.2,.7,.2}

\title{Powerful Yukawas}

\author[a,b,c]{Timothy Cohen,}
\author[a]{Matthew McCullough}
\author[a,d]{ and Neal Weiner\,}
\affiliation[a]{Theoretical Physics Department, CERN, 1211 Geneva, Switzerland}
\affiliation[b]{Theoretical Particle Physics Laboratory, EPFL, 1015 Lausanne, Switzerland}
\affiliation[c]{Institute for Fundamental Science, University of Oregon, Eugene, OR 97403, USA}
\affiliation[d]{Center for Cosmology and Particle Physics, Department of Physics,\\[2pt]
New York University, New York, NY 10003, USA}

\emailAdd{tim.cohen@cern.ch}
\emailAdd{matthew.mccullough@cern.ch}
\emailAdd{nw32@nyu.edu}

\begin{document}

\abstract{
We introduce a class of models where the masses of the light Standard Model fermions are due to an Effective Field Theory operator that appears beyond dimension-4 in the power counting expansion, resulting in a `Powerful Yukawa'.  The effective Yukawa coupling structure is UV-completed using a collective symmetry breaking pattern in the flavour sector, which we dub `Sprouted Symmetry Breaking.'  The irreducible signature is an enhanced Higgs coupling to the light Standard Model fermions.
}

\preprint{CERN-TH-2025-227}

\maketitle
\clearpage

\section{Introduction}
We often take certain aspects of nature for granted: the sun will rise, the sky is blue, the Standard Model (SM) Lagrangian includes all gauge invariant operators up to dimension 4, and so on.  In this paper, we revisit the last of these tenets, and write down a new class of phenomenologically viable models where this `fact' is not necessarily a property of the leading-order theory.

Our focus here will be on the properties of the SM matter sector.  One of the many cornerstone predictions of the SM is the 
\begin{align}
y_F = \frac{M_F}{v}
\label{eq:yFmFSM}
\end{align}
relationship between the fermion mass $M_F$ and Higgs Yukawa couplings $y_F$, where $v \simeq 174 \text{ GeV}$ is the SM Higgs vacuum expectation value (vev).  Experimentally verifying \cref{eq:yFmFSM} is often presented as a critical test of the SM picture for electroweak (EW) symmetry breaking.  For example, the universal coefficient and the linearity of this relationship between Higgs coupling and mass (mass-squared) for fermions (bosons) is often presented graphically to convey progress in LHC tests of EW symmetry breaking, see e.g.\ \cite{ATLAS:2022vkf,CMS:2022dwd}.  However, when viewing the SM as an Effective Field Theory (EFT), the constant of proportionality in \cref{eq:yFmFSM} does not reflect the fundamental structure of the SM, but rather the assumption that masses arise predominantly at lowest dimension in the SM effective theory (SMEFT).  As this paper will show, this is an assumption that may not hold if a sufficiently large symmetry, broken collectively, forbids the tree-level Yukawa.\footnote{There is a broad general literature on the possibility of modified Yukawa couplings, including \cite{Babu:1999me,Giudice:2008uua,Craig:2012vn,Dery:2013rta,Altmannshofer:2015qra,Altmannshofer:2016zrn,Chen:2017nxp,Dery:2017axi,Botella:2018gzy,Han:2021lnp,Alonso-Gonzalez:2021tpo,Goudelis:2011un,Davidson:2012ds,Dery:2014kxa,Aloni:2015wvn,Altmannshofer:2015esa,Banerjee:2016foh,Hayreter:2016aex,Buschmann:2016uzg,Zhang:2015csm,Chang:2016ave,Belusca-Maito:2016axk,Huitu:2016pwk,Barman:2022iwj,Abu-Ajamieh:2025jsz} for leptons and quarks and \cite{Bodwin:2013gca,Kagan:2014ila,Perez:2015lra,Zhou:2015wra,Brivio:2015fxa,Delaunay:2016brc,Bishara:2016jga,Soreq:2016rae,Yu:2016rvv,ATLAS:2017gko,Alves:2017avw,Coyle:2019hvs,Alasfar:2019pmn,Aguilar-Saavedra:2020rgo,Falkowski:2020znk,Egana-Ugrinovic:2021uew,Alasfar:2022vqw,Balzani:2023jas,Giannakopoulou:2024unn,Erdelyi:2024sls,Allwicher:2025mmc} with a focus on quarks alone.}

From the SMEFT point of view, a fermion mass can be generated by any operator of the form
\begin{equation}
\mathcal{O}_{\text{PY}} = c_N\frac{|H|^{2N}}{M^{2N}} \mathcal{O}_{\text{Yuk}} ~~,
\label{eq:N}
\end{equation}
where $H$ is the SM Higgs doublet, $\mathcal{O}_{\text{Yuk}}$ is the usual SM Yukawa operator, $M$ is the EFT momentum cutoff, $N$ is a non-negative integer, and $c_N$ is a dimensionless coefficient.  Since no symmetry (other than dilatation) differentiates between the operators of different values of $N$ it is typically, and moreover reasonably, assumed that the leading mass contribution arises at lowest order $N=0$.  However, much as the momentum cutoff is an integral aspect of defining an EFT, so too is the associated power counting scheme.  Both are reflections of the underlying UV completion.  The cutoff parametrizes the range of validity in terms of the EFT amplitudes, whereas the power counting reflects the spurionic structure of the UV couplings, which leaves an imprint on the IR operator field expansion.  

With this in mind, it is useful to note that the coefficient $c_N$ in \cref{eq:N} must carry at least $2 N$ more UV couplings than $\mathcal{O}_{\text{Yuk}}$. This observation reveals that the structure of couplings and symmetries in the UV could give rise to this Yukawa-like operator with $2 N$ more couplings at \emph{leading order} and, consequentially, $2 N$ more Higgs field insertions than na\"ively expected for the SM.  We call this a `Powerful Yukawa' (PY) scenario: the effective Yukawa-like coupling that gives mass to the fermions carries non-zero mass dimension.  In a PY model, the leading-order relationship between the effective Yukawa coupling $y_F^\text{PY}$ and the mass of the fermion would be 
\begin{equation}
y_F^{\text{PY}} = (2 N+1) \frac{M_F}{v} ~~,
\end{equation}
as compared to the SM prediction in \cref{eq:yFmFSM}.  The linearity is preserved, but the constant of proportionality is now set by a positive integer corresponding to the underlying `power' of the effective Yukawa.

This is not the first paper to consider this idea.  Indeed, previous authors pointed out that fermion masses that arise with the additional suppression of a factor $(v/\Lambda)^{2N}$ could be the underlying reason for the smallness of first and second generation fermion masses \cite{Babu:1999me,Giudice:2008uua}.  In contrast, the point of view we take here is not to justify or explain any mystery of the SM, but rather to ask whether the PY scenario is possible a) theoretically, in the sense that one can justify the IR power-counting as emerging from a concrete self-consistent UV scenario and b) to ask whether it is (still) possible experimentally that any of the fermion masses are \emph{dominated} by a higher dimension operator, not only in terms of Higgs Yukawa coupling and other EFT constraints, but in particular any ancillary constraints that may arise.\footnote{Note that if one were to observe a large Yukawa coupling modification that is not approximately proportional to an odd integer, this would be a sign that fermion mass is not dominated by a single higher dimension operator.}

The rest of this paper is organized as follows.  In Sec.~\ref{sec:power} we discuss the EFT generalities underpinning such a possibility.  In Sec.~\ref{sec:UV} we provide a concrete symmetry-based UV-completion and in Sec.~\ref{sec:Pheno} broad phenomenological questions are addressed before concluding.

\section{EFT for Powerful Yukawas}
\label{sec:power}
In this section, we perform a bottom-up analysis of the PY scenario.  The UV completions proposed below will rely on a collective symmetry breaking mechanism~\cite{Arkani-Hamed:2001nha,Arkani-Hamed:2002ikv}, based on a chain of U$(1)_i$ global flavor symmetries.  To simplify the presentation here, we will just consider a one-flavour SM; we will comment on more realistic extensions that address non-trivial flavour concerns in \cref{sec:Pheno} below.  We assume that the only sources of U$(1)_i$ symmetry breaking are coupling spurions $y_i$ with charge $Q_i=(1,-1)$ under $\text{U}(1)_i \times \text{U}(1)_{i+1}$, and zero charge under the rest of the U(1) symmetries.  

Let us assume that there are $2 N+1$ U(1) symmetries.  We enforce that the EFT respects this symmetry pattern, in that the only symmetry breaking is tracked by the $y_i$ spurions.  This implies that there is only one class of operator that is allowed by the symmetries at tree-level.  Up to overall numerical factors, the allowed operator takes the form
\begin{equation}
\mathcal{L}_{\text{EFT}} \supset \frac{\prod_{i=1}^{2 N} y_i}{M^{D-4}} \mathcal{O}_{\text{EFT}} ~~,
\end{equation}
where $D$ is the mass dimension of the operator $\mathcal{O}_{\text{EFT}}$, which only depends on IR fields whose combination has total charge $Q_\mathcal{O}=(-1,1)$ under $\text{U}(1)_1 \times \text{U}(1)_{2N+1}$, and $M$ is a UV mass scale.  Note that the selection rules forbid $\mathcal{O}_{\text{EFT}}$ from entering in the effective action without the additional couplings.  Suppose we identify $\text{U}(1)_1$ with $\text{U}(1)_F$, the chiral symmetry of an IR left-handed fermion, and $\text{U}(1)_{2 N+1}$ with $\text{U}(1)_{F^c}$, the chiral symmetry of an IR right-handed fermion.  Since each factor of $y_i$ implies the presence of at least one IR field in the operator $\mathcal{O}_\text{EFT}$, the lowest dimension operator that is consistent with coupling dimension,\footnote{See e.g.\ \cite{Giudice:2016yja} for one choice of $\hbar$ conventions.} gauge invariance, and the (spurionic) global symmetries has the form
\begin{equation}
\mathcal{L}_{\text{EFT}} \supset
\frac{\prod_{i=1}^{2N} y_i}{M^{2 N}} |H|^{2N} y_0 H F F^c ~~,
\label{eq:SMEFT}
\end{equation}
where $y_0$ is the spurion coupling for the relevant SM flavour symmetries.

To explore the scaling behavior of the EFT coefficient, let us assume the spurions are all equal $y_i = y_0 = y$.  After electroweak symmetry breaking, the SM fermion mass is
\begin{equation}
M_F = \frac{y}{y_T} \left(\frac{y}{y_T} \frac{M_T}{M} \right)^{2N} M_T ~~,
\end{equation}
where we have expressed this with respect to the top mass $M_T$ and top Yukawa $y_T$ to anchor the expression to a familiar SM scale.
Inverting this, we can estimate the scale where the new UV states appear for a given choice of couplings, which corresponds to the scale where the effective description breaks down.  We find
\begin{align}
M & = \frac{y}{y_T} M_T \left(\frac{y}{y_T} \frac{M_T}{M_F} \right)^{\frac{1}{2 N}} \notag\\
& \simeq 2 \text{ TeV} \times \frac{y}{4 \pi} \left(\frac{y}{4 \pi} \frac{2\text{ TeV}}{M_F}\right)^{\frac{1}{2 N}} ~~.
\end{align}
For example, for a benchmark with $N=1$ one has
\begin{align}
 M \simeq  4000 \text{ TeV} \times \bigg(\frac{y}{4 \pi} \bigg)^{\frac{3}{2}} \qquad\quad (M_F = M_\text{Electron})~~,
\end{align}
if the effective operator in \cref{eq:SMEFT} generates all of the electron mass.

The fermion Yukawa coupling is then enhanced with respect to the SM expectation by a factor
\begin{equation}
\frac{y_F}{y_{F,\text{SM}}} = 1+2 N = D-3 ~~,
\end{equation}
where $D$ is the 
dimension of the operator at which the fermion mass arises, and $y_F$ is the physical Yukawa coupling of the light fermion to the Higgs boson, and $y_{F,\text{SM}}$ is the SM Yukawa coupling.  This tells us that it could be possible for the light fermion masses, such as the up, down and electron, to arise at leading order from an operator with dimension as high as $D=16$ with the associated new states lying beyond $6$ TeV.  It is therefore plausible that new states could lie beyond the reach of HL-LHC, in a model where the light fermion Yukawa couplings is enhanced by a factor of $13$.

Note that $\mathcal{L}_{\text{EFT}}$ of \cref{eq:SMEFT} cannot be the whole story.  The reason is that no symmetry forbids lower-order operators such as
\begin{equation}
 \left(\frac{\hbar}{16 \pi^2} \frac{M^2}{|H|^2} \right)^j \mathcal{L}_{\text{EFT}} ~~,
\end{equation}
from arising.  In other words, the spurionic arguments presented above only hold at tree level.  Indeed, simply closing one pair of Higgses appearing in \cref{eq:SMEFT} into a loop yields a quadratically divergent contribution to the EFT with a coefficient $\Lambda^2/(4 \pi M)^2$, for some UV cutoff $\Lambda$. Taking $\Lambda \sim M$, as expected for a threshold correction arising at the matching scale, we find the naturally present lower dimension operators would dominate the fermion mass \emph{unless} 
\begin{equation}
 M \lesssim 4 \pi\s v ~~,
 \label{eq:PYMupperBound}
\end{equation}
in natural units.  As a result, for a UV-completion of a PY theory to be natural, the UV scale should not exceed $\sim$ few TeV, which is within the reach of the HL-LHC and future high energy colliders.

\section{A UV-Completion: Sprouting Symmetry Breaking}
\label{sec:UV}
Having now described the general features of the PY scenario, we turn to a concrete UV completion.  The generic arguments of the previous section leading to \cref{eq:PYMupperBound} imply that the upper bound on the scale of new physics must be $\lesssim$ few TeV for the effective light SM fermion Yukawa coupling from a higher dimension operator to be parametrically enhanced with respect to the natural dimension-4 contribution.  For a generic point in parameter space, $M \sim$ TeV is already ruled out by LHC searches for QCD-charged states.  We therefore present a UV-completion where all new states are only EW-charged to avoid possible conflicts with direct search limits.  It is straightforward to write down a UV model that introduces new QCD-charged states.

The model presented in this section only introduces new scalar fields that have the same quantum numbers as the SM Higgs doublet.  The structure introduced here is reminiscent of collective symmetry breaking, which was inspired by dimensional-deconstruction~\cite{Arkani-Hamed:2001kyx,Hill:2000mu}.  However, the PY UV-completion introduced here does not follow the typical pattern from a dimensional deconstruction, wherein one imagines a linear picture of adjacent nodes in theory space charged under neighbouring global symmetries.  By contrast, this model is more like a tree than a moose.  Each node `sprouts' a new branch, and the associated collective symmetry breaking structure requires introducing a two-dimensional theory space in order to understand the selection rules.  To our knowledge, such a sprouting picture for collective symmetry breaking has not been proposed before.

We start by describing one `sprout,' which can be thought of as a module that can be connected to any node.  Attaching a sprout grows the field dimension of the generated effective operator.  The sprout in this model will contain four new Higgs doublets.

We call the extra Higgses in the sprout $H_1$, $H_2$, $H_3$, $H_4$.  Higgses $H_i$ with odd-$i$ flavour indices have conjugate gauge charges to those with even-$i$ indices.  The interactions are\footnote{Note that we do not include a $H_1 \cdot H_3 ~ H_2 \cdot  H_4$ interaction because it will not contribute to the PY effective operator.}
\begin{align}
-\mathcal{L}_{\text{Int}} \supset \lambda\s H_1 \cdot H_2 ~ H_3 \cdot  H_4 + \lambda' H_1 \cdot H_4 ~ H_3 \cdot  H_2   + \sum_{i=1}^4 H_i^\ast \cdot \mathcal{O}_i +\text{h.c.}~~.
\label{eq:LIntSprout}
\end{align}
Structurally we may think of the $H_{i}$ as each carrying unit global $U(1)_i$ charge, the $\mathcal{O}_i$ as carrying unit $U(1)_i$ charge and $\lambda$, $\lambda'$ as carrying negative unit $U(1)_{1-4}$ charge.  These selection rules and gauge symmetries restrict the interactions to take the form given in \cref{eq:LIntSprout}.  In addition to these interaction terms, masses, quartics and mixed-quartics of the form $|H_i|^2 |H_{j\neq i}|^2$ are allowed.

\begin{figure}[t!]
\begin{center}
\includegraphics[width=0.4\textwidth]{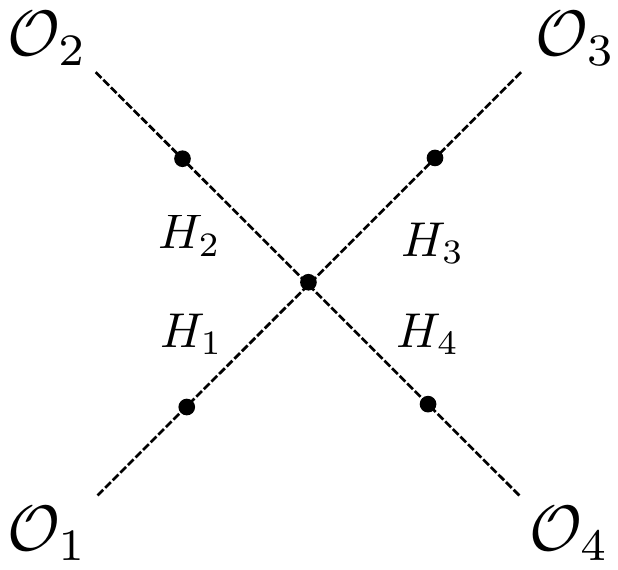} 
\caption{\label{fig:sprout} Structure of a symmetry-protected sprout.  If the operators $\mathcal{O}_{i}$ only contain light fields, this leads to an EFT operator of the form in \cref{eq:LEFT4Os}.  If any of the $\mathcal{O}_{i}$ contain heavy fields, they will be connected to further sprouts, forming a tree that leads to an even higher dimension EFT operator.} 
\end{center}
\end{figure}

Imagine that the $\mathcal{O}_i$ are built out of only SM fields.  In the region of parameter space where there is a scale separation between the $H_i$ masses and the masses of the SM fields contained within the $\mathcal{O}_i$, we can integrate out the heavy $H_i$ scalars.  This generates the tree-level term 
\begin{equation}
-\mathcal{L}_{\text{EFT}} \supset  \frac{1}{M_1^2 M_2^2 M_3^2 M_4^2} \left(\lambda\, \mathcal{O}_1 \cdot \mathcal{O}_2~ \mathcal{O}_3 \cdot \mathcal{O}_4 + \lambda'\, \mathcal{O}_1 \cdot \mathcal{O}_4 ~ \mathcal{O}_3 \cdot \mathcal{O}_2 \right) ~~.
\label{eq:LEFT4Os}
\end{equation}
The key idea of this UV completion is that the sprout is a module for which the only combination of operators involving more than one $\mathcal{O}_i$ is $\prod_{i=1}^4 \mathcal{O}_i$.  This structure then leads to the kinds of tree-level symmetry-protected combinations as depicted in \cref{fig:sprout}.  Then to generate the PY EFT structure, we identify 
\begin{subequations}
\begin{align}
\mathcal{O}_1 &= y e^c L~~, \\[3pt]
\mathcal{O}_{2,4} &= \mu^2_{2,4} H~~, \\[3pt]
\mathcal{O}_{3} &= \mu^2_{3} H^\ast~~.  
\end{align}
\end{subequations}
Following the same logic that led to \cref{eq:LEFT4Os}, this generates
\begin{equation}
-\mathcal{L}_{\text{EFT}} \supset  \frac{\lambda}{M^8} \left( \prod_{i=2}^4 \mu^2_i \right) y |H|^2 H e^c L ~~,
\label{eq:PYfromSprout}
\end{equation}
We can identify $U(1)_i$ symmetries by giving spurious charges to the $\mu^2_{2-4}$ and $e^c L$.  In the notation of \cref{eq:SMEFT}, we are identifying the product $\lambda\s \mu_2^2\s \mu_3^2\s \mu_4^2\s y$ with $y_1\s y_2\s y_0$.  This symmetry implies that (at tree-level) we generate the EFT operator in \cref{eq:PYfromSprout}, 
but we will not generate $H e^c L$ alone.

In the case that the operators $O_i$ only include light fields, the model also generates additional EFT operators of the form
\begin{equation}
\mathcal{L}_\text{EFT} \supset \frac{1}{M^2} \sum_{i=1}^4 \mathcal{O}_i^\ast \left( \frac{\Box}{M^2} \right)^n \mathcal{O}_i
\end{equation}
will also arise at tree-level, where $n\geq 0$.  Although they do not connect the various sectors, such operators may be important phenomenologically.

It is then straightforward to write down generalizations of this model that generate even higher dimension operators as the leading tree-level effect, if one or more of the $\mathcal{O}_i$ includes additional heavy fields.  This implies that they will appear as the branches of another sprout, ultimately connecting all the sprouts together to form a more complicated tree.  The extremities of the tree must end on IR fields, which allows one to generate an arbitrarily high-order operator in a symmetry-protected manner.

To see how this can be used to generate a PY EFT, consider the same $\mathcal{O}_i$ as above, but replace $\mathcal{O}_{4}$ with the $\mu^2_4 H_1'$ of an entirely new sprout, denoted by the prime.  Then, taking $\mathcal{O}'_{1,3} = (\mu_{1,3}')^2 H^\ast$ and $\mathcal{O}'_{2} = (\mu_{2}')^2 H$, one would now generate
\begin{equation}
-\mathcal{L}_{\text{EFT}} \supset  \frac{\lambda^2}{M^{16}} \left( \prod_{i=2}^4 \mu^2_i \prod_{i=1}^3 (\mu_i')^2 \right) y |H|^4 H e^c L ~~,
\end{equation}
but not $|H|^2 H e^c L$ and $H e^c L$ at tree-level.  The extension to larger branches is straightforward.

Crucially, this sprouting structure is enforced by a symmetry and thus is radiatively stable in the UV and IR:  A higher dimension operator \emph{can} dominate the Yukawa structure, even in cases where, from a purely IR perspective, lower-dimension operators would be expected to be more important.\footnote{Note that for all external legs attached to a SM Higgs fields this class of collective symmetry breaking allows for arbitrarily high-dimension operators of the form $|H|^{2 N}$ in the Higgs potential to have large coefficients.}

\subsection*{Radiative Corrections}
The possibility of generating lower-order operators in the IR theory has already been discussed in Sec.~\ref{sec:power}.  We focus on the sprouting class of UV completions and and examine the impact of radiative corrections at the matching scale.

\begin{figure}[t]
\begin{center}
\hspace{25pt}\includegraphics[width=.9\textwidth]{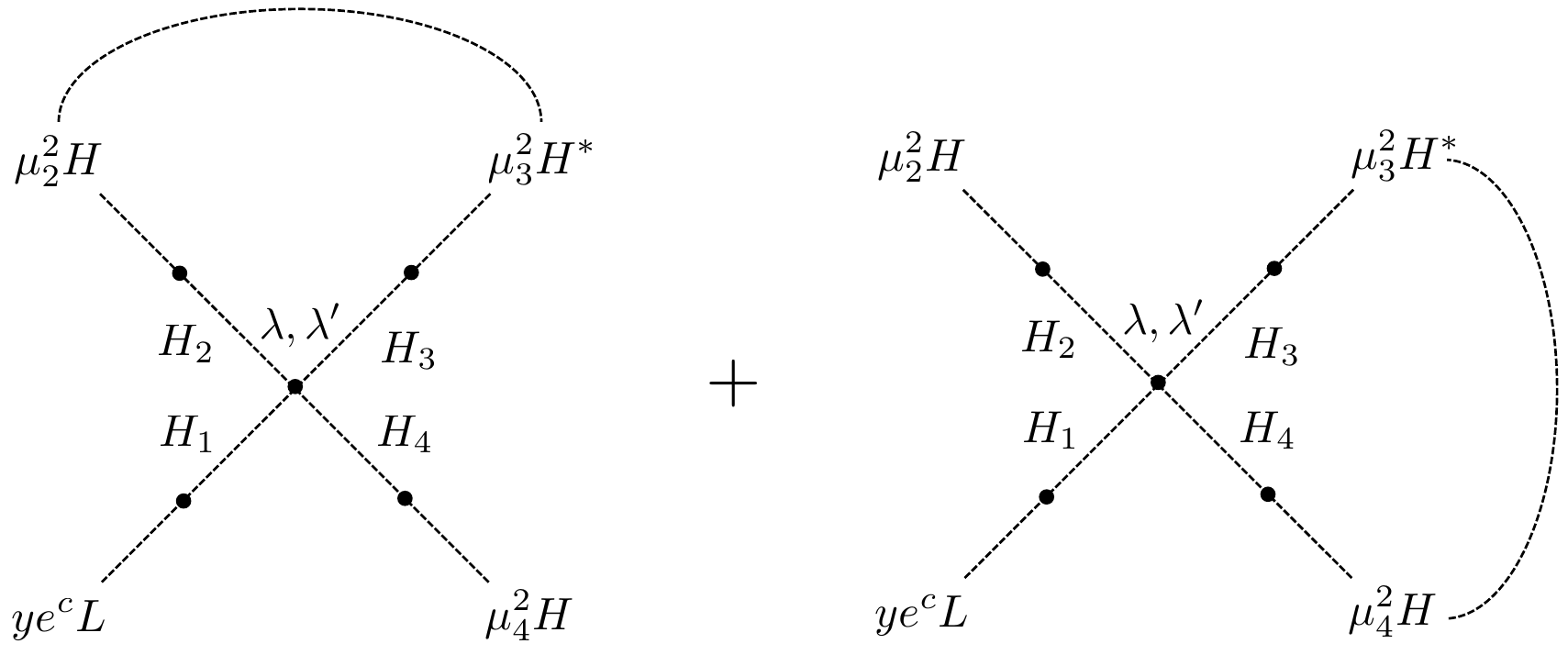} 
\caption{\label{fig:loop} One-loop diagrams that generate lower-dimension Yukawa couplings.} 
\end{center}
\end{figure}

One can explicitly calculate the correction to the lowest power ($N=0$) Yukawa from integrating out the heavier Higgs fields. In Fig.~\ref{fig:loop} we see there are two candidate Higgs field contractions, in red, which at the matching scale may generate operators with two fewer Higgs fields in the EFT.  One finds that the same structure which generates $N=1$ in the UV also gives rise to the $N=0$ operator 
\begin{equation}  
-\mathcal{L}_\text{EFT} \supset \frac{y\s  {\mu_2 }^2 {\mu_3 }^2 {\mu_4 }^2  }{8 \pi ^2 M_1^2 M_2^2 M_4^2}\left(\left(\lambda + 2 \lambda' \right) \frac{\log (M_3^2/M_2^2)}{M_3^2/M_2^2-1} + (\lambda \leftrightarrow \lambda' ,2 \leftrightarrow 4) \right)  H e^c L ~~.
\end{equation}

The na\"ive expectation of Sec.~\ref{sec:power} is confirmed.  As a result the extra Higgs doublets are expected to be heavier than the SM Higgs in order for the EFT description to be valid, but lighter than $\sim 2$ TeV in order for the powerful Yukawa scenario to be realisable. 

\section{Phenomenology}
\label{sec:Pheno}
Since the new states are only EW-charged and heavy we do not consider direct production at colliders, focussing instead on effects that can be well-described in the EFT limit.

\subsection{SMEFT Operators}
Again normalising to the top mass, assuming $\mu_i = \mu$, and that the leading contribution to a light fermion mass comes from a powerful Yukawa, one finds that 
\begin{eqnarray}
\left(\frac{\mu^2}{M^2} \right)^3 & = &  \left(\frac{y_T^2}{\lambda} \frac{M^2}{M_T^2} \right) \left(\frac{y_T}{y} \frac{M_F}{M_T} \right)^{\frac{1}{N}} 
\nonumber \\[4pt]
& \simeq & 0.2\, \frac{16 \pi^2}{\lambda} \frac{M^2}{\text{TeV}^2} \left(5 \times 10^{-7} ~\frac{4 \pi}{y} \frac{M_F}{\text{MeV}} \right)^{\frac{1}{N}}  ~~
\end{eqnarray}
in order to reproduce the observed light fermion masses.
For the lowest non-trivial case of $N=1$, one thus has $\mu^2/M^2 \gtrsim 5\times 10^{-3}$.  As a result of the sprouted symmetry selection rules, 
all SMEFT operators must be generated with at least two powers of $\mu^2/M^2$.  
We therefore expect that the effective scale suppressing any additional SMEFT operator will scale as
\begin{eqnarray}
\Lambda & \sim & \left(\frac{M^2}{\mu^2} \right) M \notag\\
& \sim & 2\s M \left(\frac{\lambda}{16 \pi^2} \frac{\text{TeV}^2}{M^2}\right)^{\frac{1}{3}} \left(2 \times 10^{6} ~\frac{y}{4 \pi} \frac{\text{MeV}}{M_F} \right)^{\frac{1}{3 N}} ~~.
\end{eqnarray}
For instance, for $N=1$ one has $\Lambda \lesssim 200\, M$ for maximised couplings, corresponding to $\Lambda \lesssim 18\s M$ for $\lambda = y^2 = 1$ in natural units.  If the couplings $y$ and $\lambda$ are small then this scale could be low enough to lead to observable consequences. However, there is ample freedom in choosing $y$ and $\lambda$ for all other SMEFT operators to be sufficiently suppressed that their observable effects are below experimental sensitivity.  The reason for this is that the effective scale associate with the operator that generates the light fermion masses in a powerful Yukawa theory is consequently very high, if it results from a theory with $\mathcal{O}(1)$ UV couplings.

Relatedly, we also expect to have, or radiatively generate, interactions of the form
\begin{equation}
-\mathcal{L}_{\text{EFT}} \supset \lambda_\mu \frac{\mu_i^2}{M^2} H \cdot H_i |H|^2
\end{equation}
in the UV theory.  However, it is plausible that $\lambda_\mu$ can be naturally a loop factor below any other quartic couplings in the UV, thus we may also neglect to consider the SMEFT operators these interactions give rise to, especially given the additional $(\mu^2/M^2)^2$ suppression which again arises.

In summary, while the sprout model does give rise to additional SMEFT operators for which sensitivity is in principle possible, the light fermion masses allow for such contributions to be small enough to be out of experimental reach.

\subsection{Flavour}
Depending on the structure of the UV, flavour could play a significant role in the phenomenology of PY models.  We will consider both leptons and quarks, for which the consequences can significantly differ. Note that these UV models can also lead to new sources of CP violation, but we are not considering these effects here.

\subsubsection*{Leptons}
The discussion here largely follows Sec. 2.3 of \cite{Allwicher:2025mmc}.  In the absence of neutrino masses the lepton sector of the SM respects a $U(1)_e\times U(1)_\mu\times U(1)_\tau$ flavour symmetry.  As a result, if the UV also respects this symmetry then any flavour-changing processes will have amplitudes proportional to neutrino masses and will hence be suppressed.  Consistent with this assumption we consider a variety of UV flavour scenarios.  Let us call `$y$' the coefficient of the renormalisable Yukawa interaction and `$\tilde{y}$' the coefficient of the operator $\mathcal{O}_{\text{PY}}$.
\begin{itemize}
\item Electrophilic: $\tilde{y} \propto$ diag(0,0,1) in the fermion mass basis. In this case, if the sprouts only couple to the electron then the only ancillary constraint is through the electron anomalous magnetic moment, which does not strongly constrain this scenario.
\item  MFV-like $\tilde{y} \propto y$:  Constraints on $\tau$ Yukawa couplings already rule out a PY scenario for the electron.
\item  Universal  $\tilde{y} \propto \mathbb{I}_3$:  Significant electron Yukawa corrections are possible and consistent with current bounds.  
\item  Breaking from the $U(1)_e\times U(1)_\mu\times U(1)_\tau$ assumption, in an anarchic flavour scenario, constraints from $\mu \to e \gamma$ \cite{MEGII:2025gzr} are in marginal tension with an $N=2$ PY scenario for the electron.  Higher choices of $N$ would be in significantly greater tension.
\end{itemize}
The theoretical plausibility of these various scenarios differs.  However, the conclusion that universal or anarchic flavour structures can be consistent with a powerful Yukawa is suggestive that, at least for the electron, the overall possibility appears to be phenomenologically viable.

\subsubsection*{Quarks}
For general discussions of enhanced light-quark Yukawas, see \cite{Egana-Ugrinovic:2019dqu,Giannakopoulou:2024unn}.  In principle, if the light generation quark masses result from PYs, this would enhance multi-boson production \cite{Egana-Ugrinovic:2021uew}.  However, for the factor $\sim$ few enhancements considered here this effect would be below experimental sensitivity.  From a flavour perspective, the situation is somewhat different for quarks as compared to leptons since FCNC processes are already possible in the SM, albeit not at tree-level.    As a result, engineering of $u,d$-philic, MFV-like and Universal scenarios may be significantly more challenging than for leptons.
\begin{itemize}
\item $u,d$-philic: $\tilde{y} \propto$ diag(0,0,1) in the fermion mass basis. In this case, if the sprouts only couple to a first generation quark then Higgs couplings provide the leading constraint.
\item  MFV-like $\tilde{y} \propto y$:  Constraints on $t,b$ Yukawa couplings already rule out a PY scenario for the light quarks.
\item  Universal  $\tilde{y} \propto \mathbb{I}_3$:  Significant light quark Yukawa corrections are possible and consistent with current bounds.  
\item  In an anarchic flavour scenario, constraints from meson oscillations, in particular Kaon oscillations, are particularly strong \cite{Harnik:2012pb}.  Therefore, anarchic flavour is inconsistent with a PY scenario for the light quarks.
\end{itemize}

\section{Conclusions}
In this paper, we revisited the standard lore that the dynamics of the Standard Model fields are described, at leading order, by the renormalizable interactions of the Standard Model.  
The smallness of the light fermion Yukawa couplings leaves open the possibility that the light fermions fundamentally have stronger interactions with the Higgs, but that their small mass is due to these interactions arising from interactions with new heavy states.  Such a scenario is described at low energies through non-renormalisable SM interactions and is characterised by a very non-SM-like Higgs Yukawa coupling.

While such a non-minimal scenario is not forced upon us by any data, it remains possible as an explanation of at least the lightest generation masses. But perhaps more importantly, this highlights the importance of keeping in mind the varieties of underlying UV power counting when working with the SMEFT, illustrated through the class of `sprout' models introduced here.  It is interesting to ask if similar lessons could hold in other areas of the SM.  For instance, could significant sources of custodial symmetry violation lurk at nearby energy scales, hidden by the fact that the symmetry-breaking does not feed into the dimenion-6 SMEFT at leading order?

This work also raises a broader theoretical question.  Dispersion relations show that in the EFT momentum expansion, lower-dimension contributions cannot be arbitrarily suppressed relative to their higher-dimension counterparts, see, for instance, App A. of \cite{Englert:2019zmt}.  The present work suggests that similar lessons do not hold for the field expansion, where generically a Wilson coefficient hierarchy of a loop factor may be tolerable.  It would be interesting to ask if there are fundamental theoretical constraints on how large such a hierarchy can naturally be.

Of broader theoretical interest is the Sprouting Symmetry Breaking class of collective symmetry breaking models presented here.  Through collective symmetry breaking, this class of models can motivate hierarchies in the field expansion, where Wilson coefficients grow in magnitude as one includes operators with more fields.  Structurally, these models are interesting as they do not follow the pattern expected for classic collective symmetry breaking constructions based on dimensional deconstruction.  Rather, the nodal structure is more reminiscent of a web, or branch.  It would be interesting to see if alternative applications may be found for this obscure class of models.

The Standard Model is mature.  However the equation printed on t-shirts and \href{https://visit.cern/content/standard-model-mug}{mugs}  cannot yet be taken for granted.  It remains possible that the light fermions receive most of their mass from an operator beyond the conventions of the Standard Model.

\acknowledgments
T.C.\ is supported by the DOE under grant DE-SC0011640.
N.W.\ is supported by the NSF under grant PHY-2210498, by the BSF under grant 2018140, and by the Simons Foundation.

\clearpage

\bibliographystyle{JHEP}
\bibliography{main}

\end{document}